\documentclass[a4paper]{jpconf}
\usepackage{graphicx}

\def\apj{{\it Astrophys.~J.}}

\def\prd{{\it Phys.~Rev.}~D}

\def\mnras{{\it Mon.~Not. Roy.~Astr.~Soc.}}

\def\grg{{\it Gen. Rel. Grav.}}

\def\jcap{{\it J. Cosmol. Astropart. Phys.}}

\begin{document}
\title{Observational cosmology and the cosmic distance-duality relation}

\author{S Jhingan$^1$, D Jain$^2$ and R Nair$^1$, }

\address{$^1$ Centre for Theoretical Physics, Jamia Millia
Islamia, New Delhi 110025, India}
\address{$^2$ Deen Dayal Upadhyaya College, University of Delhi, New Delhi 110015, India}

\ead{sanjay.jhingan@gmail.com}

\begin{abstract}
We study the validity of cosmic distance duality relation between
angular diameter and luminosity distances. To test this duality
relation we use the latest Union2 Supernovae Type Ia (SNe Ia) data
for estimating the luminosity distance. The estimation of angular
diameter distance comes from the samples of galaxy clusters (real
and mock) and FRIIb radio galaxies. We parameterize the distance
duality relation as a function of redshift in four different ways
and we find that the mock data set, which assumes a spherical
isothermal $\beta$ model for the galaxy clusters does not
accommodate the distance duality relation while the real data set
which assumes elliptical $\beta$ model does.
\end{abstract}

\section{Introduction}
The reciprocity relation relates the distances between two events
(say  the source $S$ and the observer $O$), which are connected by
null geodesics. This relation is of fundamental importance in
observational cosmology especially measurements such as distant type
Ia supernovae (SNe Ia), cosmic microwave background (CMB), galaxy
observations and gravitational lensing \cite{ellisgrg}. Bassett $\&$
Kunz {\cite{bk}} explored whether this relation can shed some light
on the presence of exotic physics and they ruled out
non-accelerating models of universe (replenishing dust model) by
more than $4\sigma$ level.
\begin{figure}[h]
\begin{minipage}{8pc}
\includegraphics[width=8pc,height=8pc]{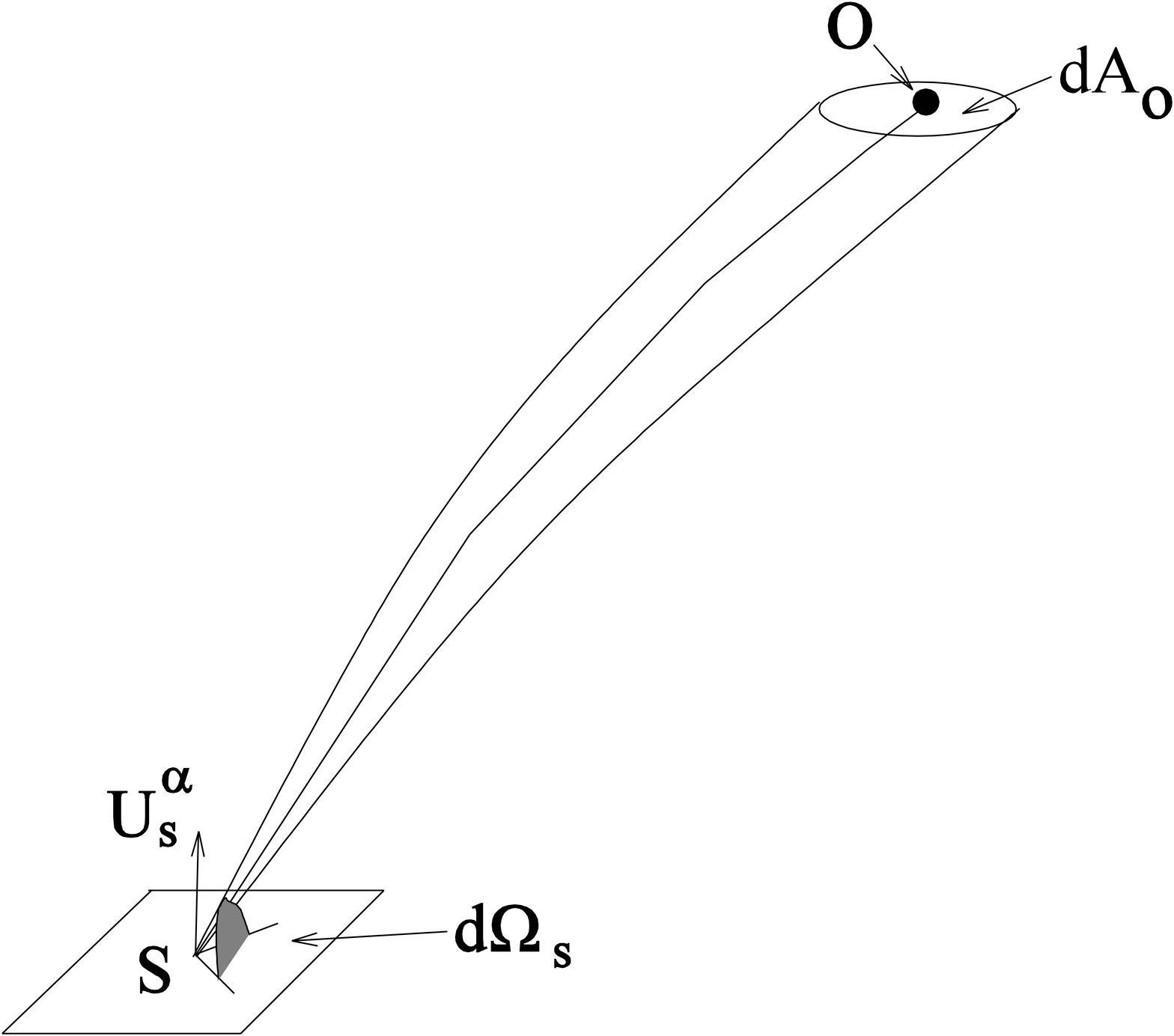}
\end{minipage}\hspace{3pc}
\begin{minipage}{8pc}
\includegraphics[width=8pc,height=8pc]{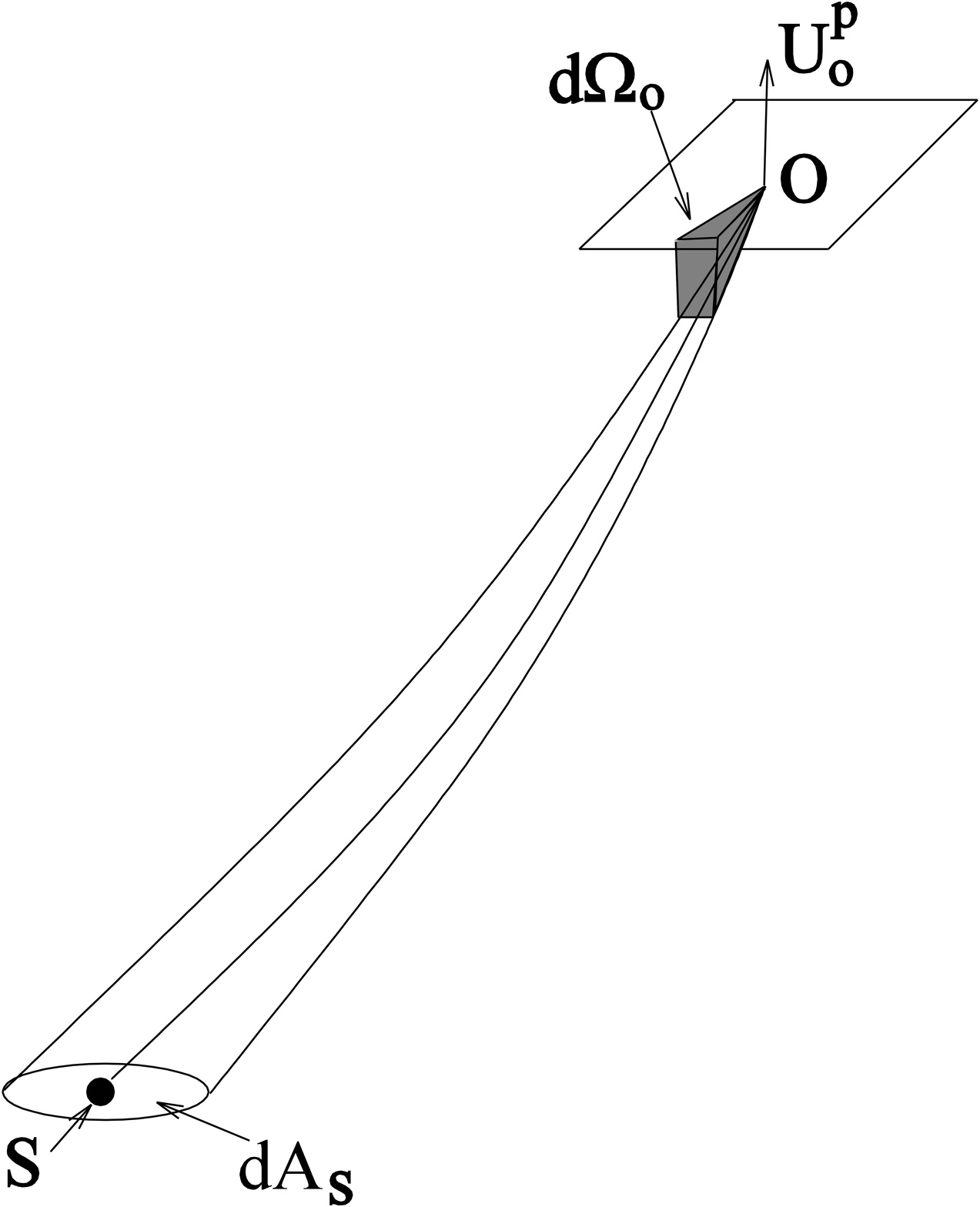}\hspace{3pc}
\end{minipage}
\begin{minipage}{18pc}\hspace{5pc} \caption{\small{Consider a bundle of null geodesics
emanating from the source subtending a solid angle $d\Omega_s$. This
bundle has a cross section $dA$, and the source angular distance
$D(U^\alpha _s)$ is defined as $D(U^\alpha _s) =
\left(\frac{dA}{d\Omega _s}\right)^\frac{1}{2}$. Similarly the
observer area distance $D(U^p _s)$ is defined as $D(U^p _s) =
\left(\frac{dA_s}{d\Omega _o}\right)^\frac{1}{2}$.} }\label{new}
\end{minipage}
\end{figure}

\subsection*{The reciprocity relation}
If the geodesic deviation equation holds and photon travel on null
geodesics, it can be  shown that (see figure \ref{new}), $D(U^\alpha
_s)$ and $D(U^p _s)$ are related as \cite{dd,ehl}
\begin{equation}
D(U^\alpha _s) = (1+z) D(U^p _s)\label{reci}
\end{equation}
This is known as the reciprocity relation. Note that this relation
holds  regardless of the metric or the matter content of the
space-time. Using the reciprocity relation and assuming that the
total number of photons are conserved on the  cosmic scales (above
mentioned observational methods are all based on this fundamental
assumption), one can derive a relation between the angular diameter
distance and the luminosity distance \cite{ehl}:
\begin{equation}
d_L=d_A (1+z)^2
\end{equation}

This equation is known as the $distance~ duality$ relation (DD) and
it is testable by astronomical observations if $d_L$ and $d_A$ are
known for a source. To study the validity of this relation, we
analyse the following red-shift dependence of DD
\begin{equation}\label{ddrelation}
\eta(z) \equiv \frac{{d_L}}{d_A(1+z)^2 } \;.
\end{equation}

Uzan et al.\cite{uz}, used the combined measurements of
Sunyaev-Zeldovich effect and  X-ray emission data of galaxy
clusters and showed that if DD relation does not hold then the angular diameter
distance measured from the clusters is $d_A^{cluster}(z)= d_A\,
\eta^2$ , and hence the DD relation (\ref{ddrelation}) gets modified to
\begin{equation}\label{erro}
\eta(z) \equiv \frac{{d_A^{cluster}(1+z)^2}}{d_L}  \,.
\end{equation}

In this work our aim is to reanalyze the validity of DD relation in
a comprehensive manner by using different data samples and
parameterizations \cite{us}.
\section{Parameterizations, data and method}

\subsection{$\eta$(z) parameterizations}
Our parameterizations are inspired by model independent
parameterizations for the dark energy equation of state \cite{jo}.
We are parameterizing $\eta(z)$ whose value stays one when photon
number is conserved, gravity is described by a metric theory and
photons travel on null geodesics. Any significant violation from the
DD relation will hint at the  breakdown of one or more of these
assumptions. To model any departure from unity we parameterize
$\eta$ with four parametric representations:

\begin{center}
$\eta_{I}(z) =  \eta_{0}+\eta_{1} z $ \\
$~~\eta_{II}(z) = \eta_{2} + \eta_{3}  \frac{z}{1+z}$ \\
$~~~~\eta_{III}(z) = \eta_{4} + \eta_{5} \frac{ z}{(1+z)^2}$\\
$~~~~~~~~\eta_{IV}(z) = \eta_{6}-\eta_{7} \ln(1+z)$\\
\end{center}

\subsection {Data}
For the luminosity distances, we use latest
Union2 SNe Ia data \cite{am}. We consider that pair of galaxy
cluster/radio galaxy and SNe for which $\Delta z < 0.005$ \cite{h1}.
Because of this condition, the number of data points are limited to
24, 222 and 12 for the data set I, II and III respectively.

\begin{itemize}

\item Data Set I: This sample consists of 25
galaxy clusters (isothermal elliptical $\beta$ model, with concordance model
for the cosmological distance-redshift relationship) \cite{fi}. The redshift interval for clusters
in  this sample is $0.02 < z < 0.78$.
\item Data Set II: This data set contains 578 angular diameter distances of mock clusters. The redshift distribution of this sample
is from $0.05 < z < 0.76$. For more details see the ref.'s
\cite{sa1,sa2}.
\item Data Set III: In this sample the calculation of angular diameter
distance is obtained by using the physical size of extended radio
galaxies \cite{ru}. This data set contains 20 radio galaxies up to
redshift z = 1.8.

\end{itemize}

\subsection{Method}
We perform the $\chi^2$ analysis to fit the parameters of the
assumed parameterizations.
\begin{equation}
\chi^2(p) = \sum_{i}\frac{({\eta_{th}(z_i,p)} -
{\eta_{obs}(z_i))^2}} { {\sigma(z_i)}^2 } \; .
\end{equation}
Where $\eta_{th}$ is the assumed form of parameterizations and $\eta_{obs}$ is the observed value of
$\eta$, which is calculated by using $d_L$ and $d_A$ at a particular value
of redshift. The
unknown parameters are denoted by the
variable $p$.

The $\eta_{obs}$ for the radio galaxies are obtained by assuming  $d_A^{radio} =
d_A$ (R. A. Daly, private communication) in eq. (\ref{ddrelation}). The
angular diameter distance for radio galaxies is given in terms of the
dimensionless coordinate
distance, $y$, as
\begin{equation}
d_{A}(i) = \frac{y(z_i)}{H_0 (1 + z_i)} \; ,
\end{equation}
where  $y = a_{0} r H_0 $, and $a_0 r$ is the coordinate distance
\cite{ru}. Treating Hubble constant, $H_0$, as a nuisance parameter
we marginalize over it by assuming gaussian prior. In this analysis distances are obtained by assuming the flat
$\Lambda$-CDM universe.

\section{Results and Discussions}

We have studied the validity of DD relation using different  data
sets, assuming four general parameterizations of $\eta(z)$,
completely in analogy with the varying equation of state of dark
energy, $\omega(z)$. Our results are summarized as follows:
\begin{itemize}

\item Table \ref{tab:datasetI}, \ref{tab:datasetII} and
\ref{tab:datasetIII} show the best fit values of the parameters for
the four parameterizations for the first, second and third data set
respectively.

\begin{center}
\begin{table}[h]
  \caption{Best fit values for all parameterizations - data set I}
  \vspace{0.2cm}
  \label{tab:datasetI}
  \centering
  \begin{tabular}{c c c}
    \hline
  { $\chi^2_{\nu}$} & {Parameters} & {Parameters} \\
    \hline
1.217 \quad & $\eta_{0} = 0.999 \pm 0.144$ \quad&
 $\eta_{1}=
 -0.058 \pm 0.507$ \quad\\
1.216 \quad&  $\eta_{2} = 1.007 \pm 0.170$ \quad&
 $\eta_{3} = -0.118 \pm 0.822 $
 \quad\\
1.216 \quad& $\eta_{4} = 1.013 \pm 0.205$ \quad&
 $\eta_{5} =
 -0.198 \pm 1.345$ \quad  \\
1.216 & $\eta_{6} = 1.003 \pm 0.157$ \quad& $\eta_{7} =   ~~0.087 \pm 0.651$ \\\hline
    \hline
  \end{tabular}
\end{table}
\end{center}

\begin{center}
\begin{table}[h]
  \caption{Best fit values for all parameterizations - data set
  II } \vspace{0.2cm}
  \label{tab:datasetII}
  \centering
  \begin{tabular}{c c c}
    \hline
   {\bf $\chi^2_{\nu}$} & {Parameters} & {Parameters} \\
    \hline
  1.076 & $\eta_{0} = 0.973 \pm 0.048$ & $\eta_{1} =  -0.108
\pm 0.159$ \\
1.078 & $\eta_{2} = 0.971 \pm 0.058$ & $\eta_{3} = -0.138
\pm 0.267$ \\
1.080 & $\eta_{4} = 0.972 \pm 0.072 $ & $\eta_{5} =
-0.141 \pm 0.453$ \\
1.077 & $\eta_{6} = 0.972 \pm 0.053$ & $\eta_{7} =
~~0.124 \pm 0.208$ \\\hline
    \hline
  \end{tabular}
\end{table}
\end{center}

\begin{figure}[h]
\begin{minipage}{14pc}\hspace{2pc}
\includegraphics[width=14pc]{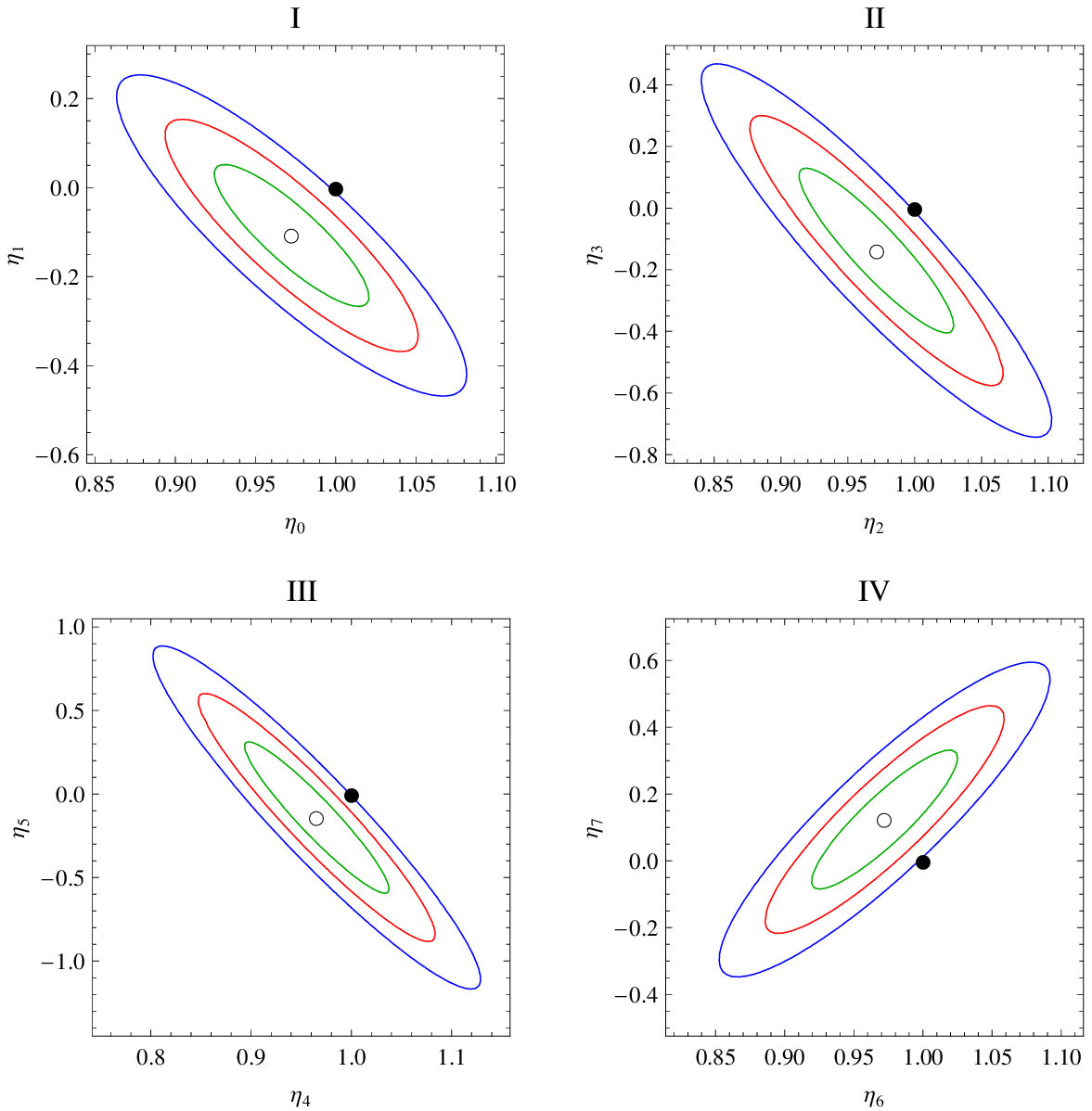}
\end{minipage}\hspace{5pc}
\begin{minipage}{14pc}
\includegraphics[width=14pc]{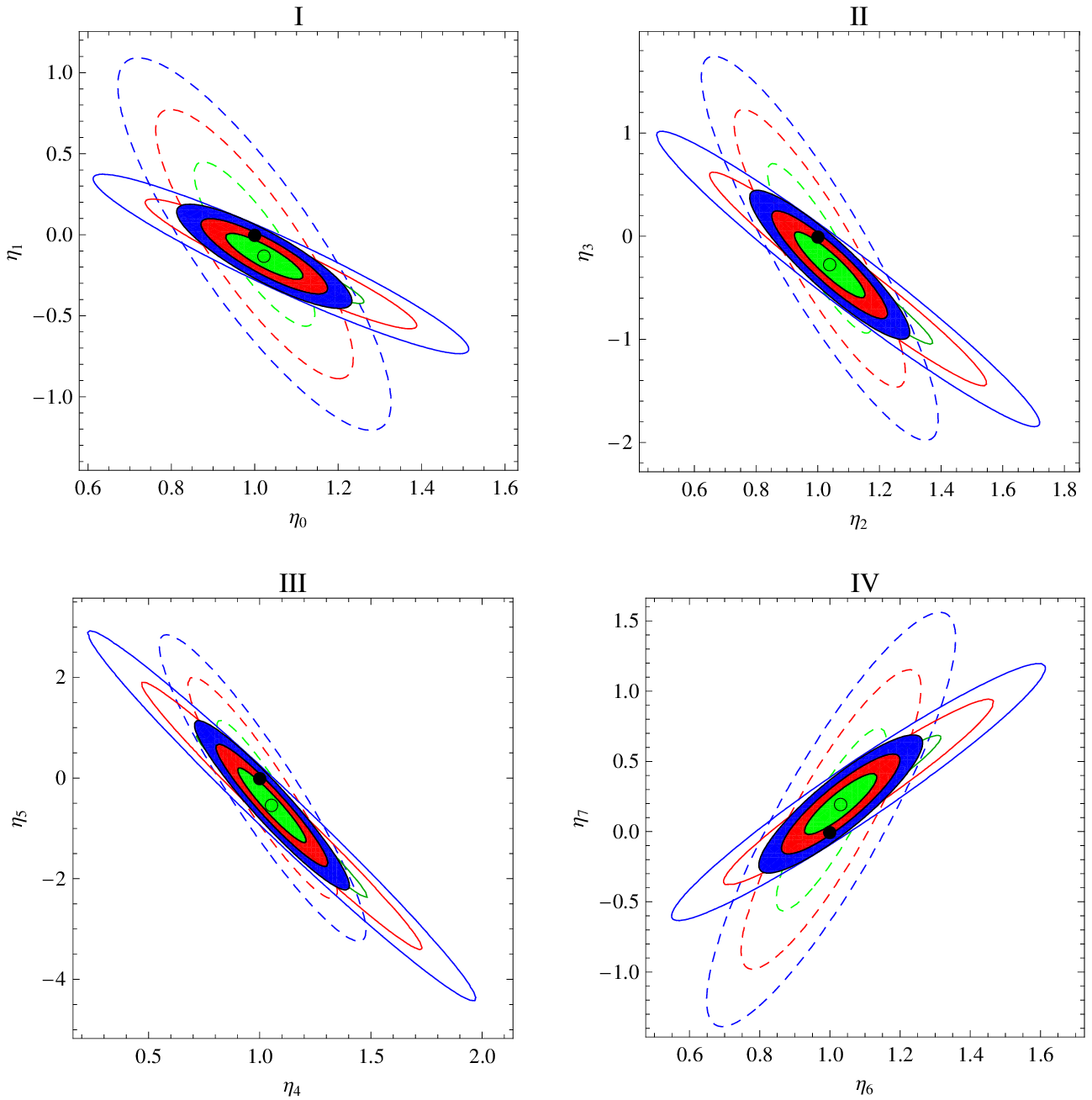}
\end{minipage}
\caption{\small{The left panels shows $1\sigma$, $2\sigma$ and
$3\sigma$  contours in $\eta_i -\eta_j$ plane with data set II. In
the right panel, dashed, solid and filled contours correspond to
data set I, III and (I + III) respectively. In both the figures, the
position of filled circle in the contours indicate the point where
$\eta(z=0)=1$ and the position of empty circle indicates the best
fit value of the parameters.}}\label{allfig}
\end{figure}

\begin{table}[h]
  \caption{Best fit values for all parameterizations - data set
  III} \vspace{0.2cm}
  \label{tab:datasetIII}
  \centering
  \begin{tabular}{c c c}
    \br
   {\bf $\chi^2_{\nu}$} & {Parameters} & {Parameters} \\
    \mr
    0.944 & $\eta_{0} = 1.063 \pm 0.198 $ & $\eta_{1} = -0.180 \pm 0.244$  \\
0.971 & $\eta_{2} = 1.099 \pm 0.274 $ & $\eta_{3} = -0.415 \pm 0.632$ \\
1.021 & $\eta_{4} = 1.099 \pm 0.385 $ & $\eta_{5} = -0.749 \pm 1.624$\\
0.958 & $\eta_{6} = 1.081 \pm 0.235 $ & $\eta_{7} = ~ 0.282 \pm 0.404~~$\\
    \br
  \end{tabular}
\end{table}


\item All the parameterizations show degenerate behaviour with the given data
sets. The parameterizations are in complete concordance with DD
relation within 1$\sigma$ level for data set I and within 2$\sigma $
level for data set III. As expected, the bigger data set of mock
galaxy clusters (data set II), which is generated by assuming  the
spherical isothermal $\beta$  model for clusters gives tighter
constraints on various parameterizations, but most of the
parameterizations show significant deviation from DD relation (see
figure \ref{allfig}). But the real galaxy cluster data (data set I)
which is obtained by assuming isothermal elliptical $\beta$ model
for clusters is in good agreement with DD relation.

\end{itemize}

\end{document}